\documentclass{JHEP3}

\usepackage[dvips]{graphicx}
\usepackage{cite}
\usepackage{amsfonts,bbm,amsmath}
\usepackage{amssymb,amsthm}

\newcommand{\sand}[3]{\langle#1|#2|#3 \rangle}
\newcommand{\rd}{{\rm d}}

\def\IM {\mbox{Im}\,}
\def\RE {\mbox{Re}\,}

\def\beq {\begin{equation}}
\def\eeq {\end{equation}}
\def\bea {\begin{eqnarray}}
\def\eea {\end{eqnarray}}

\def\nn {\nonumber}

\def\bfl {\mbox{\boldmath $\lambda$}}

\def\tauKpi{\tau\to K \pi\nu_\tau}

\preprint{UAB--FT--682}

\title{\boldmath{$K\pi$} vector form factor constrained  by \boldmath{$\tauKpi$} and \boldmath{$K_{l_{3}}$} decays}

\author{D.~R.~Boito\\
Grup de F\'{\i}sica Te\`orica and IFAE, Universitat Aut\`onoma de Barcelona,\\
E-08193 Bellaterra (Barcelona), Spain\\
Email:~\email{boito@ifae.es}
}

\author{R.~Escribano\\
Grup de F\'{\i}sica Te\`orica and IFAE, Universitat Aut\`onoma de Barcelona,\\
E-08193 Bellaterra (Barcelona), Spain\\
Email:~\email{rescriba@ifae.es}
}

\author{M.~Jamin\\
Instituci\'o Catalana de Recerca i Estudis Avan\c cats (ICREA),\\
IFAE and Grup de F\'{\i}sica Te\`orica, Universitat Aut\`onoma de
Barcelona,\\
E-08193 Bellaterra (Barcelona), Spain\\
Email:~\email{jamin@ifae.es}
}

\abstract{
Dispersive representations of the $K\pi$ vector and scalar form
factors are used to fit the spectrum of $\tauKpi$ obtained by the Belle
collaboration incorporating constraints from results for
$K_{l_{3}}$ decays.  The slope and curvature of the vector form factor
are obtained directly from the data through the use of a
three-times-subtracted dispersion relation. We find
$\lambda_+'=(25.49 \pm 0.31) \times 10^{-3}$ and $\lambda_+''= (12.22
\pm 0.14) \times 10^{-4}$.  From the pole position on the second
Riemann sheet the mass and width of the $K^*(892)^{\pm}$ are found to
be $m_{K^*(892)^\pm}=892.0\pm 0.5$~MeV and $\Gamma_{K^*(892)^\pm}=46.5\pm 1.1$~MeV.  
The phase-space integrals needed for $K_{l_3}$ decays are calculated as well. Furthermore, the $K\pi$ isospin-$1/2$ $P$-wave 
threshold parameters are derived from
the phase of the vector form factor. For the scattering length and the effective range we find
respectively  $a_{1}^{1/2}\,= (
0.166\pm 0.004)\,m_\pi^{-3}$ and $b_{1}^{1/2}\,=( 0.258\pm
0.009)\,m_\pi^{-5}$.}

%Use showkeys class option if keyword display desired

\begin{document}

\section{Introduction}
\label{intro}

The differential decay distributions of $K\to \pi\, l\, \nu_l$ ($K_{l_{3}}$)
and $\tauKpi$ decays are governed by two Lorentz-invariant $K\pi$ form
factors that encode the non-perturbative physics, namely the vector,
denoted $F^{K\pi}_+(q^2)$, and the scalar,
$F^{K\pi}_0(q^2)$. According to the kinematical configuration, $q$
represents the exchanged ($K_{l_3}$) or the total ($\tauKpi$) $K\pi$
four-momentum. A good knowledge of these form factors is of
fundamental importance for the determination of many parameters of the
Standard Model, such as the quark-mixing matrix element $|V_{us}|$
obtained from $K_{l_{3}}$ decays~\cite{LR84}, or the strange-quark
mass $m_s$ determined from the scalar QCD strange spectral
function~\cite{JOP2}. Recently, several collaborations have produced
data for $K_{l_{3}}$ decays and new high-statistics data for $\tauKpi$
have been published by the $B$ factories. The new data sets provide
the substrate for up-to-date theoretical analyses of the $K\pi$ form
factors.

Historically, the main source of experimental information on $K\pi$
form factors have been  $K_{l_{3}}$ decays. Recently, five
experiments have collected data on semileptonic and leptonic $K$
decays: BNL-E865~\cite{E865gen}, KLOE~\cite{KLOEgen},
KTeV~\cite{KTeVgen}, ISTRA+~\cite{ISTRAgen}, and
NA48~\cite{NA48gen}. The results from these analyses yielded
an important amount of information on form factors as well as
stringent tests of QCD at low-energies and of the Standard Model
itself (for recent reviews on theoretical and experimental aspects of
kaon physics we refer to Refs.~\cite{FlaviaNet, Antonelli10}).  
Additional knowledge on the $K\pi$ form factors can be gained from
the dominant Cabibbo-suppressed $\tau$ decay: the channel $\tauKpi$.
The $\tau$ is the only
known lepton heavy enough to decay into hadrons and its hadronic decays
constitute a rather clean environment for the study of QCD at
relatively low energies~\cite{alphas} and notably for the determination of
the QCD coupling $\alpha_s$~\cite{Baikov:2008jh,ddhmz08,bj08,mal08}.  
In the 1990s, the $K\pi$ spectrum for
$\tauKpi$  was measured by ALEPH~\cite{aleph99} and
OPAL~\cite{opal04}. Lately, however, the $B$ factories have become a superior
source of high-statistics data for this reaction by virtue of the
important cross-section for $e^+e^-\to \tau^+\tau^-$ around the
$\Upsilon(4S)$ peak. As a result, as many as $10^9$ $\tau$ pairs were
recorded by Belle and BaBar~\cite{tauBfac}. A detailed spectrum for
$\tau \to K_S\,\pi^-\nu_\tau$ produced and analysed by Belle was published in
2008~\cite{Belle} with an  event sample larger than in the LEP
experiments by almost a factor of 65, allowing for a detailed analysis
of its shape. Also, a preliminary BaBar spectrum with similar
statistics has appeared recently in conference
proceedings~\cite{Babar} and, finally, BESIII should produce results
for this decay in the future~\cite{BESIII}.

On the theory side, a salient feature of the form factors in
 the kinematical region relevant for $K_{l_3}$ decays, {\emph i.e.}
 $m_l^2<q^2<(m_K-m_\pi)^2$, is that they are real.  Within the allowed
 phase-space they admit a Taylor expansion and the energy
 dependence is customarily translated into constants
 $\lambda_{+,0}^{(n)}$ defined as\footnote{From now on we refrain from writing the superscript
$K\pi$ on the form factors.}
\beq
F_{+,0}(q^2)=F_{+,0}(0)\left[1 + \lambda_{+,0}' \frac{q^2}{m_{\pi^-}^2} + \frac{1}{2}\lambda_{+,0}'' \left( \frac{q^2}{m_{\pi^-}^2}\right)^2 + \cdots \right]\,.
\label{FFTaylor}
\eeq
In $\tauKpi$ decays, however, since $(m_K+m_\pi)^2<q^2<m_\tau^2$, one
deals with a different kinematical regime in which the form factors
develop imaginary parts, rendering the expansion of
Eq.~(\ref{FFTaylor}) inadmissible.  One must then resort to more
sophisticated treatments. Moreover, in order to fully benefit from the available
experimental data, it is desirable to employ representations of the form factors
that are valid for both $K_{l_{3}}$ and $\tauKpi$ decays. 
% In the
%original Belle analysis~\cite{Belle} a weighted sum of
%Breit-Wigner-like propagators was used in order to account for the
%effect of resonances.  
In Ref.~\cite{JPP2006}, a new expression for
$F_+(s)$ was derived within the Resonance Chiral Theory (RChT)
framework~\cite{RChT} and, subsequently, the authors reanalysed the
Belle spectrum for the decay $\tauKpi$ with success~\cite{JPP2008}.
This analysis yielded new values for the constants $\lambda_{+}' $ and
$\lambda_{+}''$ emphasising the interplay between $\tauKpi$ and
$K_{l_{3}}$ experiments.

 From general principles of analyticity, the form factors must fulfil
 a dispersion relation. Unitarity provides an additional constraint on
 the imaginary part of the form factors, rendering possible the design
 of dispersive representations of $F_+$ and $F_0$ that are suited to
 describe both $\tauKpi$ and $K_{l_{3}}$ decays.  For the vector form
 factor, a step towards this feat was taken in Refs.~\cite{BEJ,BEJ2}
 where we introduced several subtracted dispersive representations of
 $F_+$. Our final proposal was a three-times-subtracted dispersive
 representation in which $\lambda_{+}'$ and $\lambda_{+}''$ are
 parameters that were determined via a  successful fit to the
 Belle spectrum. A similar dispersive approach to $F_+$ was presented
 in Ref.~\cite{Bernard:2009zm} and has been used by the KTeV
 collaboration to fit their $K_{l_{3}}$ spectra~\cite{KTeV}. Finally,
 a dispersive representation for $F_+$ that includes inelastic effects
 was introduced in Ref.~\cite{Moussallam}.  Concerning the scalar form
 factor, a thorough description that takes into account analyticity,
 unitarity, the large-$N_c$ limit of QCD, and the coupling to $K\eta$
 and $K\eta'$ channels was introduced in Ref.~\cite{JOP1} and updated
 in Refs.~\cite{JOP2,JOP3,JOP4}.  Another single-channel dispersive
 representation of $F_0$ can be found in Ref.~\cite{Bernardetal} and
a description based on the so-called method of unitarity bounds was
recently presented in Ref.~\cite{Anant}.

The main purpose of our paper is to produce an analysis of the Belle
spectrum for $\tauKpi$ incorporating constraints from experimental
results on $K_{l_{3}}$ decays. We have already advocated that such a
combined treatment of both reactions could further our knowledge of the
form factors hence paving the way for a better determination of
$|V_{us}|$~\cite{BEJ2}. Moreover, we aim at extracting as much
information as possible from the $\tauKpi$ spectrum. With the present
statistics the spectrum allows for a study of $K\pi$ dynamics in the
$P$ wave, which gives the prevailing contribution to the
decay. Watson's theorem~\cite{Watson} guarantees that below inelastic
thresholds the phase of the form factor equals the scattering phase and, therefore,
one can perform a  study of the dominant $K\pi$ $P$-wave threshold parameters. 
In addition, it has been shown~\cite{JPP2008,BEJ} that the present statistics
permits a competitive determination of the pole position of the
$K^*(892)^{\pm}$ as well as the position of a second vector resonance,
although less precisely in the latter case. Here, we determine these
two poles exploiting a novel strategy in which  fits are
done directly in terms of the physical pole positions on the second
Riemann sheet. This improvement with respect to previous
works~\cite{Belle,JPP2008,BEJ} yields a  determination of the
pole positions with a better control of uncertainties and correlations.

 In our analysis, for the vector form factor we employ the dispersive
 representation of Ref.~\cite{BEJ} whereas for the scalar $K\pi$ form
 factor we use the up-to-date results of Ref.~\cite{JOP4}. Since the
 details of these descriptions can be found in the original works,
 here we shall concentrate on the results that arise from our fit,
 namely {\it i)} the pole positions for the $K^*(892)^\pm$ and 
 $K^*(1410)^\pm$ resonances, {\it ii)} $\lambda_+'$ and $\lambda_+''$,
 {\it iii)} the result of the phase-space integrals needed in
 $K_{l_3}$ decays, and {\it iv)} the $K\pi$ isospin-$1/2$ $P$-wave
 scattering phase and the respective threshold parameters.

Our paper is organised as follows. First, in Sec.~\ref{DispFF}, we
briefly review the dispersive treatment of the vector and scalar
$K\pi$ form factors. Then, in Sec.~\ref{Fit1}, we present a fit to
$\tauKpi$ data alone. In Sec.~\ref{Fit2}, the results for a fit
incorporating constraints from $K_{l_{3}}$ experiments are given and,
in Sec.~\ref{IntKl3}, we derive our results for the phase-space
integrals relevant for $K_{l_3}$ experiments. We discuss the results
for the $K\pi$ threshold parameters and scattering phase shifts in
Sec.~\ref{Res}. Our final results and a comparison with other results
found in the literature are presented in Sec.~\ref{Conc}.

%%%%%%%%%%%%%%% SEC II: THE FORM FACTORS %%%%%%%%%%%%%
\section{Dispersive \boldmath{$K\pi$} form factors }
\label{DispFF}

The $K\pi$ form factors are defined as follows~\cite{FlaviaNet}
\beq
\sand{\pi^-(p)}{\bar s\, \gamma^{\,\mu} \,u}{K^0(k)}=       \left[  (k+p)^\mu  - \frac{m_K^2 -m_\pi^2}{q^2}(k-p)^\mu\right]F_+(q^2)   + \frac{m_K^2 -m_\pi^2}{q^2}(k-p)^\mu F_0(q^2)\,  , \label{FF}
\eeq
where $F_+(q^2)$ and $F_{0}(q^2)$ are the vector and scalar form
factors respectively and $ q^2 = (k-p)^2$. It follows from the
definition that  both form factors share the same normalisation at zero $F_+(0)= F_0(0)$.  For convenience, we
work with normalised form factors $\tilde F_{+,0}(q^2)$ such that
\beq
\tilde F_{+,0}(q^2)\equiv\frac{F_{+,0}(q^2)}{ F_{+}(0)}\,.
\eeq
First, in determinations of $|V_{us}|$, a reliable value for the
normalisation at zero is crucial in order to disentangle the product
$|V_{us}|F_{+}(0)$. In this respect, Chiral Perturbation Theory (ChPT) and
lattice QCD are the most trustworthy methods to obtain
$F_{+}(0)$. Here we are concerned with another aspect of the form
factors, namely their  energy dependence encoded in $\tilde
F_{+,0}(q^2)$. The precise knowledge of $\tilde F_{+,0}(q^2)$ is
needed when performing the phase space integrals for $K_{l_{3}}$ decays
or when studying in detail the $\tauKpi$ spectrum.  Finally, one
should bear in mind that when considering $\tau$ decays, one deals
with a crossing-symmetric version of Eq.~(\ref{FF}) for the $K\pi$
pair is in the final state.  In this case, $q^2\equiv s =
(k+p)^2>(m_K+m_\pi)^2$ and the form factors develop imaginary parts.

In  $\tauKpi$ and $K_{e_{3}}$ decays, the term containing
the vector form factor $F_+(q^2)$  dominates the differential
decay widths. The form factor, in its turn, receives a prevailing
contribution from the $K^*(892)$. This fact motivated the description
of Refs.~\cite{JPP2006,JPP2008} within RChT, which was based on an
analogous treatment of the pion vector form
factor~\cite{SCandPich,GuerreroPich97}. Although dominated by the $K^*(892)$, the authors
of Refs.~\cite{JPP2006,JPP2008} noted that a second resonance,
identified with the $K^*(1410)$, must be included in  $F_+(s)$ to
account for the higher-energy part of the $\tauKpi$ spectrum. The
description of Refs.~\cite{JPP2006,JPP2008}, albeit successful, has a slight
drawback, namely it satisfies the analyticity constraints only in a
perturbative sense. Although the violation of analyticity is expected to
be of higher orders in the chiral expansion, a description based on a
dispersive treatment was necessary to corroborate this pattern. In
Ref.~\cite{BEJ} we designed such dispersive representations of $F_+(s)$.

The rationale for our approach is as follows. From general principles, the form factor must satisfy a dispersion
relation. Supplementing this constraint with unitarity, the dispersion
relation has a well-known closed-form solution within the elastic
approximation referred to as the Omn\`es
representation~\cite{Omnes}. Although simple, this solution requires
the detailed knowledge of the phase of $F_+(s)$ up to infinity, which
 is unrealistic. An advantageous strategy to circumvent this
problem is the use of additional  subtractions, as done for the pion form factor
in Ref.~\cite{PP2001}. Subtractions in the dispersion relation entail a
suppression of the integrand in the dispersion integral for higher
energies. An $n$-times-subtracted form factor exhibits a suppression
of $s^{-(n+1)}$ in the integrand. Thereby, the information that was
previously contained in the high-energy part of the integral is
translated into $n-1$ subtraction constants. In Ref.~\cite{BEJ} we
performed fits to the Belle spectrum of $\tauKpi$ varying the number
of subtractions and testing the description with one and two vector
resonances. The outcome of these tests, described in detail in
Ref.~\cite{BEJ}, is that for our purposes an optimal description of
$F_+(s)$ was reached with three subtractions and two resonances. Here
we quote the resulting  expression
\beq
\tilde F_+(s) \,=\, \exp\left [ \alpha_1\, \frac{s}{m_{\pi^-}^2} +
\frac{1}{2}\alpha_2\frac{s^2}{m_{\pi^-}^4}   + \frac{s^3}{\!\pi}
\int\limits^{s_{\rm cut}}_{s_{K\pi}} \!\!ds'\, \frac{\delta(s')}
{(s')^3(s'-s-i0) }\right] \,.\label{dispFF}
\eeq
In the last equation, $s_{K\pi}=(m_{K^0}+m_{\pi^-})^2$ and the two subtraction constants $\alpha_1$ and
$\alpha_2$ are related to the Taylor expansion of Eq.~(\ref{FFTaylor})
as $\lambda_+'=\alpha_1$ and $\lambda_+''=\alpha_2+\alpha_1^2$.  It is
opportune to treat them as free parameters that capture our ignorance
of the higher energy part of the integral.  The constants
$\lambda_+'$ and $\lambda_+''$ can then be determined through the
fit. The main advantage of this procedure, advocated for example in Refs.~\cite{PP2001,Bernardetal,Bernard:2009zm,BEJ}, 
is that the subtraction constants turn out to be less model dependent
as they are determined by the best fit to the data. The calculation of these
constants, on the other hand, depends strongly on the perfect knowledge of $\delta(s)$.
However, since now $\alpha_{1,2}$ are determined by the data, in the 
limit $s\to \infty$ the asymptotic behaviour of $F_+(s)$ cannot be satisfied. This is so because
a perfect cancellation between terms containing $\alpha_{1}$ and $\alpha_{2}$
with polynomial terms coming from the dispersion integral must occur in order
to guarantee that $F_+(s)$ vanishes  as $1/s$. We have checked that our form factor, within the entire
range where we apply it (and beyond), is indeed a decreasing function of $s$ which 
renders this approach credible.

With Eq.~(\ref{dispFF}), the transition from the kinematical
region of $\tauKpi$ to that of $K_{l_{3}}$ decays is straightforward and
the dominant low-energy behaviour of $F_+(s)$ is encoded in
$\lambda_+'$ and $\lambda_+''$. The cut-off $s_{\rm cut}$ in the
dispersion integral is introduced to quantify the suppression of the
higher energy part of the integrand. The stability of the results is
checked varying this cut-off in a wide range from $1.8\, \mbox{GeV} <
\sqrt{s_{\rm cut}} < \infty$.
 It is important to stress that Eq.~(\ref{dispFF}) remains valid
beyond the elastic approximation provided $\delta(s)$ is the
phase of the form factor, instead of the corresponding scattering phase. But, of course, in order to  employ it in practice  we must 
 have a model for the phase. As described in detail
 in Ref.~\cite{BEJ}, we take a form inspired by the RChT treatment
 of~Refs.~\cite{JPP2006,JPP2008} with two vector resonances.  Here we
 relegate the details concerning $\delta$ to
 Appendix~\ref{AppA}. However, one important remark is in order.
 Since we keep the real part of the loop bubble integral $\RE \tilde H(s)$ in
 Eq.~(\ref{Den}), the mass and width parameters of Ref.~\cite{BEJ} are
 shifted as compared with those of
 Refs.~\cite{Belle,JPP2006,JPP2008}. This shift emphasises the need
 for the computation of the {\it physical} pole position of the
 resonances on the second Riemann sheet. We have shown~\cite{BEJ,BEJ2}
 that although the mass and width parameters from
 Refs.~\cite{Belle,JPP2008,BEJ} differ considerably, the pole
 positions arising from the models are in good agreement. To clarify this issue further, in this work we implement a
 numerical improvement in our codes that allows us to perform the fits
 directly in terms of the pole positions on the second Riemann
 sheet. This new procedure is clearer as it avoids the cumbersome
 intermediate stage where one must compute the pole positions from the
 unphysical parameters to obtain meaningful
 results~\cite{Escribanoetal}. Furthermore, correlations and
 uncertainties are obtained directly for the physical poles and are
 therefore  more reliable.

In the previous analysis of Refs.~\cite{JPP2008,BEJ} the scalar form
factor was shown to play an important role for the low-energy part of
the $\tauKpi$ spectrum, between threshold and $\sim 0.8$~GeV. On the other hand, the fit was not very
sensitive to the details of $F_0$ as it is in the case of
$F_+$. Therefore, we again rely on the coupled channel representation of
$F_0$ first presented in Ref.~\cite{JOP1} and updated in
Refs.~\cite{JOP2,JOP3,JOP4}. The main features of this treatment can be
found in Appendix~\ref{AppA}.

%%%%%%%%%%%% SEC III: FITS TO TAU---> K PI NU ONLY %%%%%%%%%%%%%%%%
\section{Fit to \boldmath{$\tauKpi$}}
\label{Fit1}

Before proceeding to a fit that combines information from $\tauKpi$
and $K_{l_{3}}$ data, we shall perform in this section a short update of
Ref.~\cite{BEJ}. The aim is twofold. First we want to ascertain the
impact of performing the fit directly in terms of the physical pole
positions for the vector resonances. Second, the results of this
section  serve as a point of reference for the new analysis.  For the sake
of completeness, we recall here how the $K\pi$ form factors enter the
description of $\tauKpi$. 

Assuming isospin invariance, the differential decay
distribution for $\tau\to K\pi \nu_\tau$ can be cast in terms of the
$K\pi$ form factors as
\bea
 \frac{\rd\Gamma_{K\pi}}{\rd\sqrt{s}} =&& \frac{G_F^2|V_{us}F_{+}(0)|^2 m_\tau^3}
{32\pi^3s}\,S_{\mbox{\tiny EW}}\left(1-\frac{s}{m_\tau^2}\right)^{\!2}     \times \nn \\
 &&\times \left[
\left( 1+2\,\frac{s}{m_\tau^2}\right) q_{K\pi}^3\,|\tilde F_+(s)|^2 +
\frac{3\Delta_{K\pi}^2}{4s}\,q_{K\pi}|\tilde F_0(s)|^2 \right] \, , 
\label{dGamma}
\eea
where we summed over the two possible decay channels $\tau^-\to \bar K^0
\pi^- \nu_\tau$ and $\tau^-\to  K^- \pi^0 \nu_\tau$ that contribute
in the ratio $2:1$. In Eq.~(\ref{dGamma}), $S_{\mbox{\tiny EW}}$ is an
electroweak correction factor, $\Delta_{K\pi}\equiv m_K^2-m_\pi^2$,
$s=(k+p)^2$ with $k$ and $p$ being respectively the kaon and pion
momenta, and
\beq
\label{qKpi}
q_{K\pi}(s) \,=\, \frac{1}{2\sqrt{s}}\sqrt{\Big(s-(m_K+
m_\pi)^2\Big) \Big(s-(m_K-m_\pi)^2\Big)} \times \theta\Big(s-(m_K+m_\pi)^2\Big) 
\eeq
 is the kaon momentum in the rest frame of the hadronic
system.  In order to analyse the data, one must rely on an ansatz for the number of
events observed in a given bin of the experimental spectrum. As explained in Ref.~\cite{JPP2008} the theoretical
number of events $N_i^{\rm th}$ in the $i$-th bin is taken to be
\beq
\label{Nth}
N_i^{\rm th} =\mathcal{N}_T\,  \frac{1}{2}\,\frac{2}{3}  \,\Delta^{i}_{\rm b}\,  \frac{1}{\Gamma_\tau \, \bar B_{K\pi}}  \frac{\rd\Gamma_{K\pi}}{\rd\sqrt{s}} (s_{\rm b}^ i)\,,
\eeq
where $\mathcal{N}_T$ is the  total number of events, the factor $\frac{1}{2}$ and $\frac{2}{3}$ account
for the fact that the $K_S\pi^-$ channel was analysed, $\Delta^{i}_{\rm b}$ is the  width of the $i$-th bin, $\Gamma_\tau$  is the  total $\tau$ decay  width, $\bar
B_{K\pi}$ is a normalisation  constant that, for a perfect description
of the spectrum, should be the $\tauKpi$ branching ratio, and, finally,
$s_{\rm b}^ i $ is the centre of the $i$-th bin. In the case of Belle's
spectrum~\cite{Belle} one has $\mathcal{N}_T=53110$ and a constant bin width $\Delta_{\rm b}=11.5$~MeV.

In this fit, we minimise the $\chi^2$ function given by
\beq
\label{chi2tau}
\chi^2 = \sum_{i=1}^{90}{}^{\prime} \left(\frac{N_i^{\rm th} - N^{\rm exp}_i}{\sigma_{N^{\rm exp}_i}} \right )^2  + \left(\frac {\bar B_{K\pi} - B_{K\pi}^{\rm exp}}{\sigma_{B^{\rm exp}_{K\pi}}}\right)^2\,,
\eeq
where $N^{\rm exp}_i$ and $\sigma_{N^{\rm exp}_i}$ are, respectively,
the experimental number of events and the corresponding uncertainty in the $i$-th
bin. The prime in the symbol of sum indicates that bins 5, 6, and 7
are excluded from the minimisation\footnote{If these three points are
  included in the fit the results do not change significantly although
  the $\chi^2$ is  larger. Furthermore, there is no indication for a
  peak at this energy and BaBar spectra do not display a bump close to
  threshold. Hence, we decided, following Refs.~\cite{JPP2008, BEJ},
  to exclude these points. For a visual account,  points not included
in the $\chi^2$ are shown as unfilled circles in Fig.~\ref{MainFitFigure}.}.
In the $\chi^2$, following a suggestion of the experimentalists~\cite{Epifanov}, 
we include data up to  bin number 90 which corresponds
to $\sqrt{s}=1.65925$~GeV. Finally, the lowest data point is not
taken into account since, with physical meson masses, its centre lies below 
the $K\pi$ threshold.
 The second term on the right-hand side of Eq.~(\ref{chi2tau}) was not
included in the $\chi^2$ function of Ref.~\cite{BEJ}.  It introduces
an additional restriction that allows us to treat the normalisation
$\bar B_{K\pi}$ of Eq.~(\ref{Nth}) as a free parameter.
%The theoretical value $B_{K\pi}^{th}$ is obtained
%from the integration of Eq.~(\ref{dGamma}) and, In a good fit, the
%numbers $\bar B_{K\pi}$ and $B_{K\pi}^{\rm exp}$ should be compatible.
Then, the parameters of the fit are 8 in total. First, the two
constants $\lambda_+'$ and $\lambda_+''$ responsible for the behaviour
of $\tilde F_+(s)$ near the origin. Second, the five parameters that
determine the resonance properties, {\it i.e.} the complex pole
positions of the $K^*(892)$ and\footnote{For simplicity, in tables we refer to the  $K^*(892)$ and the $K^*(1410)$ simply as $K^*$ and $K^{*\prime}$ respectively.} $K^*(1410)$ and the mixing parameter
$\gamma$ [see Eq.~(\ref{FpKpi2})].  The mass and width of the
resonances are extracted from the complex pole position $s_R$
as~\cite{Escribanoetal}
\beq
\label{pole}
\sqrt{s_R}=m_R - \frac{i}{2} \Gamma_R\,.
\eeq
The phase of the form factor is fully determined by the latter set of
parameters.  The 8th parameter of the fit is the normalisation $\bar
B_{K\pi}$.

 In the fit, we employ the following numerical values: $|V_{us}|F_+(0)=
 0.2163(5)$~\cite{Antonelli10}, $G_F=1.16637(1)\times 10
 ^{-5}$~GeV$^{-2}$~\cite{PDG}, $m_\tau=1776.84$~MeV~\cite{PDG},
 $S_{\rm EW}=1.0201(3)$~\cite{ErlerSew},
 $f_\pi=92.21(14)$~MeV~\cite{Rosner}, $f_K/f_\pi=
 1.197(6)$~\cite{Rosner}, and $B^{\rm
   exp}_{K\pi}=0.418(11)\%$~\cite{BKpi1,BKpi2}.  We recall that the cut-off
 $s_{\rm cut}$ of Eq.~(\ref{dispFF}) has to be varied in order to
 check the stability of the results upon the high-energy part of the
 dispersion integral. When quoting final results one must therefore
 include an uncertainty due to the small residual dependence on
 $s_{\rm cut}$.  The results of fits with four values of $s_{\rm
   cut}$, namely $s_{\rm cut}=3.24$ GeV$^2$, 4 GeV$^2$, 9 GeV$^2$, and
 $s_{\rm cut}\to \infty$, are displayed in Tab.~\ref{fitTauKpi}.

%\begin{widetext}
\TABLE{
%\begin{center}
%\begin{tabular}[!ht]
%\begin{ruledtabular}
\begin{tabular} {|l c c c c|}
\hline\hline
                       & $s_{\rm cut}=3.24$ GeV$^2$& $s_{\rm cut}=4$ GeV$^2$  & $s_{\rm cut}=9$ GeV$^2$   &$s_{\rm cut}\to \infty$ \\ 
\hline
$\bar B_{K\pi}$         &  $0.416\pm 0.011\%$  &     $0.417 \pm 0.011\%  $  & $0.418\pm0.011 \%$ & $0.418\pm0.011\%$ \\
$(B^{\rm th}_{K\pi})$    &  $(0.414\%)$         &      $(0.414\%)$        &     $(0.415\%)$       &    $(0.415\%)$    \\
 $m_{K^*}$ [MeV]        & $892.00 \pm 0.19$    &    $892.02 \pm 0.19$    &   $892.03 \pm 0.19$   &  $892.03 \pm 0.19$\\
 $\Gamma_{K^*}$  [MeV]  &   $46.14 \pm 0.44$   &     $46.20 \pm 0.43$    &    $46.25 \pm 0.42$   &  $46.25 \pm 0.42$ \\
 $m_{K^{*\prime}}$  [MeV]     & $1281^{+25}_{-33} $   &     $1280^{+25}_{-28} $  &    $1278^{+26}_{-27} $ & $1278^{+26}_{-27}$ \\
 $\Gamma_{K^{*\prime}}$  [MeV]&     $243^{+92}_{-70}$ &      $193^{+72}_{-56}$   &  $177^{+66}_{-52}$     &  $177^{+66}_{-52}$ \\
$\gamma \times 10^2$   &  $-5.1^{+1.7}_{-2.6}$  & $-3.9^{+1.3}_{-1.8}$    &   $-3.4^{+1.1}_{-1.6}$& $-3.4^{+1.1}_{-1.6}$\\
$\lambda_+^{'} \times 10^{3}$ &  $24.15 \pm 0.72$ & $24.55 \pm 0.68$    & $24.86 \pm 0.66$     & $24.88 \pm 0.66$   \\
$\lambda_+^{''}\times 10^{4}$ & $11.99 \pm 0.19 $& $11.95 \pm 0.19$    & $11.93  \pm 0.19 $    &$11.93 \pm 0.19$  \\
$\chi^2/{\rm n.d.f.}$         & 74.1/79           & 75.7/79               & 77.2/79              & 77.3/79            \\
\hline\hline\end{tabular}
\caption{Results for the fit to Belle's  $\tauKpi$ spectrum~\cite{Belle}.  As a consistency check, for each one of the fits we give the value  $B^{\rm th}_{K\pi}$ obtained from the integration of  Eq.~(\ref{dGamma}).   }\label{fitTauKpi}}
%\end{ruledtabular}
%\end{table}
%\end{center}
%\end{widetext}

Some of the results of Tab.~\ref{fitTauKpi} are to be compared with
those of Tab.~4.2 of Ref.~\cite{BEJ}.  Concerning $\lambda_+'$,
$\lambda_+''$ and $\gamma$ they are very similar if not identical.
However, in Ref.~\cite{BEJ}, the $\chi^2$ that was minimised did not
include the second term in the right-hand side of Eq.~(\ref{chi2tau})
and therefore $\bar B_{K\pi}$ was kept fixed or, otherwise, the strong
positive correlation between $\bar B_{K\pi}$ and the constant
$\lambda_+'$ would render a good determination of these parameters impracticable.  The main
difference between the two fits lies, as already stressed, in the pole
positions of the vector resonances. In Tab.~\ref{fitTauKpi}, the
results correspond to physical masses and widths obtained from the
pole positions in the second Riemann sheet according to
Eq.~(\ref{pole}) whereas the results of Tab.~4.2 of Ref.~\cite{BEJ}
are non-physical parameters. Consequently, results for masses and
widths presented here should not be directly compared with the
parameters of Tab.~4.2 of Ref.~\cite{BEJ}. Instead, one should compare
with the physical poles that can be found in Eqs.~(5.4) and (5.5) of
Ref.~\cite{BEJ}. One sees that  the central values agree nicely.   On the other hand,
the treatment of the uncertainties affecting the poles is here much
more trustworthy. The results of Tab.~\ref{fitTauKpi} come from an
analysis performed by the MINOS function of the CERN-Minuit
library. The errors are smaller than the ones quoted in
Ref.~\cite{BEJ} due to the proper inclusion of correlations. 
Finally, the fit is very
stable against changes in $s_{\rm cut}$. This is specially true for
the mass and width of the $K^*(892)$ but in all other cases variations
are at most at the level of one standard deviation. 
In order to produce a feeling for  the correlation coefficients between the
parameters of our fits, as an example we display in Tab.~\ref{Correlations1} those corresponding to
$s_{\rm cut}=4$~GeV$^{2}$.

\TABLE{
\begin{tabular} {|l c c c c c c c c|}
\hline\hline
 & $\bar B_{K\pi}$ &  $m_{K^*}$ & $\Gamma_{K^*}$  &$m_{K^{*\prime}}$ &  $\Gamma_{K^{*\prime}}$  &$\gamma $  &  $\lambda_+^{'} $   &  $\lambda_+^{''}$      \\ 
\hline
%$\bar B_{K\pi}$& 1  &  &  &  &  &  &  &  \\
$m_{K^*}$& -0.119 & 1  &  &  &  &  &  &  \\
$\Gamma_{K^*}$& 0.041 & -0.017 & 1  &  &  &  &  &  \\
$m_{K^{*\prime}}$ &-0.048 & -0.168 & -0.158 & 1  &  &  &  &  \\
 $\Gamma_{K^{*\prime}}$&0.110& 0.182 & 0.303 & -0.628 & 1  &  &  &  \\
 $\gamma $ &-0.148 & -0.244 & -0.425 & 0.558 & -0.865 & 1  &  &  \\
 $\lambda_+^{'} $&0.711 & 0.008 & 0.543 & -0.298 & 0.462 & -0.653 & 1  &  \\
$\lambda_+^{''}$ & 0.880 & -0.132 & 0.421 & -0.212 & 0.355 & -0.466 & 0.934 & 1\\
\hline\hline
\end{tabular}
\caption{Correlation coefficients for the parameters of the fit with  $s_{\rm cut}=4$~GeV$^{2}$, third column of Tab.~\ref{fitTauKpi}.}
\label{Correlations1}
}

%%%%%%%%%%%% SEC III: FITS WITH CONSTRAINTS FROM K--> pi e nu  %%%%%%%%%%%%%%%%

\section{\boldmath\rm\bf Fit to \boldmath{$\tauKpi$} with restrictions from $K_{l_3}$}
\label{Fit2}

 Dispersive representations of $K\pi$ form factors can be used in
 order to  simultaneously fit both $\tauKpi$ and $K_{l_{3}}$
 spectra. We have recently advocated~\cite{BEJ2}, performing Monte
 Carlo (MC) simulations, that the main benefit of such a combined fit
 would be the reduction of the uncertainties on the parameters
 $\lambda_+'$ and $\lambda_+''$, leading to smaller uncertainties
 in the phase-space integrals needed for the extraction of $V_{us}$ from
 kaon decays. For the want of an unfolded data set from $K_{l_{3}}$
 experiments, we perform here a fit to $\tauKpi$ constrained by
 results for $\lambda_+'$ and $\lambda_+''$ obtained from a
 compilation of $K_{l_{3}}$ analyses~\cite{Antonelli10}.

In results  obtained
from quadratic representations such as the one of
Eq.~(\ref{FFTaylor}), the errors on $\lambda_{+,0}^{(n)}$ have a clear
statistical meaning. In principle, therefore, it is straightforward to
include that information in the $\chi^2$ that is to be  minimised by the
fit. In this case, the statistical correlation between $\lambda_+'$ and 
$\lambda_+''$ must be taken into account.
The $\chi^2$ to be minimised contains then one additional term
 \beq
\label{chi2Comb}
 \chi^2 = \sum_{i=1}^{90}{}^{\prime} \left(\frac{N_i^{\rm th} - N^{\rm exp}_i}{\sigma_{N^{\rm exp}_i}} \right )^2  + \left(\frac {\bar B_{K\pi} - B_{K\pi}^{\rm exp}}{\sigma_{B^{\rm exp}_{K\pi}}}\right)^2
+  (\bfl_+^{\rm th} -\bfl_+^{\rm exp})^{\rm T} V^{-1}  (\bfl_+^{\rm th} -\bfl_+^{\rm exp})\, ,
\eeq
where the first two terms in the right-hand side are the same as in
Eq.~(\ref{chi2tau}) whereas the last one encodes the information
from $K_{l_{3}}$ analyses. In this last term, the vectors $\bfl_ +^{\rm th, exp}$ are given by
\beq
\bfl_ +^{\rm th, exp}=
\left(\begin{array}{c}
\lambda_+^{\prime\, \rm th, exp} \\
\lambda_+^{\prime\prime\, \rm th, exp}
\end{array}\right)\, ,
\eeq
and the $2\times 2$  matrix $V$ is the experimental error matrix for $\bfl_+$  such that
\beq
V_{ij}= \rho_{ij}\, \sigma_i\, \sigma_ j\, ,
\eeq
where the indices  refer to $\lambda_+'$ and $\lambda_+''$,
$\rho_{ij}$ is the correlation coefficient ($\rho_{ij}=1$ if $i=j$),
and $\sigma_{i}$ the experimental errors on $\lambda_+'$ and
$\lambda_+''$.  For the experimental values we employ the results of
the compilation of $K_L$ analyses performed by Antonelli et al. for the FlaviaNet Working Group on Kaon Decays in
Ref.~\cite{Antonelli10}: $\lambda_+^{\prime\, \rm exp}=(24.9\pm1.1)\times 10^{-3}$, 
$\lambda_+^{\prime\prime\, \rm exp}=(16\pm 5)\times 10^{-4}$ and
$\rho_{\lambda_+^{\prime},\lambda_+^{\prime \prime}}=-0.95$.  Results
for  fits using the $\chi^2$ function of Eq.~(\ref{chi2Comb}) with $s_{\rm cut}=3.24$ GeV$^2$, 4 GeV$^2$, 9 GeV$^2$,
and $s_{\rm cut}\to \infty$ are shown in Tab.~\ref{fitCombined}. In Fig.~\ref{MainFitFigure}, 
the Belle spectrum for $\tauKpi$ is confronted with the results for the fit with $s_{\rm cut}=4$~GeV$^2$.
Finally, as an example, the correlation matrix for $s_{\rm cut}=4$~GeV$^2$ is given in Tab.~\ref{Correlations2}.

\TABLE{
\begin{tabular} {|l c c c c|}
\hline\hline
  &  $s_{\rm cut}=3.24$ GeV$^2$  &  $s_{\rm cut}=4$ GeV$^2$ & $s_{\rm cut}=9$ GeV$^2$  & $s_{\rm cut}\to \infty$  \\ 
\hline
$B_{K\pi}$        &      $ 0.429 \pm 0.009$       &    $0.427 \pm 0.008  \%   $        &   $0.426 \pm 0.008 \%$   & $0.426\pm 0.008 \%$\\
$(B^{\rm th}_{K\pi})$            &  $(0.426\%)$         &      $(0.425\%)$        &     $(0.423\%)$       &    $(0.423\%)$    \\
 $m_{K^*}$ [MeV]       &      $892.04 \pm 0.20$        &    $892.02 \pm 0.20$       &   $892.03 \pm 0.19$   & $892.03 \pm 0.19$\\
 $\Gamma_{K^*}$  [MeV]       &      $46.58 \pm 0.38$        &    $46.52 \pm 0.38$       &   $46.48 \pm 0.38$   & $46.48 \pm 0.38$\\
 $m_{K^{*\prime}}$  [MeV]       &       $1257^{+30}_{-45} $        &    $1268 ^{+25}_{-32}$        &   $1270^{+24}_{-29}$  & $1271^{+24}_{-29}$\\
 $\Gamma_{K^{*\prime}}$  [MeV]       &      $321^{+95}_{-76} $        &     $238^{+75}_{-57}$       &   $206 ^{+67}_{-50}$   &  $205^{+67}_{-50}$\\
$\gamma \times 10^2$ &       $-8.2^{+2.2}_{-3.5} $         &    $-5.4 ^{+1.4}_{-2.0}$         &    $-4.4 ^{+1.2}_{-1.6}$    &  $-4.4^{+1.2}_{-1.6}$ \\
 $\lambda_+^{'} \times 10^{3}$&      $25.43 \pm 0.30$        &    $25.49 \pm 0.30$        &   $25.55 \pm 0.30$    & $25.55 \pm 0.30$ \\
$\lambda_+^{''}\times 10^{4}$&       $12.31 \pm 0.10$         &    $12.20 \pm 0.10$        &   $12.12 \pm 0.10$     &  $12.12\pm 0.10$  \\
$\chi^2/{\rm n.d.f.}$ &      77.9/81       &    78.1 /81       &   79.0 /81  & 79.1/81\\
\hline\hline
 \end{tabular}\caption{Results of fits to Belle spectrum~\cite{Belle} of $\tauKpi$ with constraints from the $K_{l_3}$ analysis of Ref.~\cite{Antonelli10}. The $\chi^2$ function is defined in Eq.~(\ref{chi2Comb}). As a consistency check, for each one of the fits we give the value  $B^{\rm th}_{K\pi}$ obtained from the integration of  Eq.~(\ref{dGamma}).} 
\label{fitCombined}}

Comparing the results of the fit constrained by $K_{l_3}$ analyses,
Tab.~\ref{fitCombined}, with the results of the fit to $\tauKpi$
alone, Tab.~\ref{fitTauKpi}, one sees that the statistical uncertainty
in $\lambda_+'$ and $\lambda_+''$ is reduced roughly by a factor of
2. Another advantage of the new fit is that the results for
$\lambda_+'$ and $\lambda_+''$ are much more stable against changes in
$s_{\rm cut}$. The errors in $\lambda_+'$ are  largely dominated by statistics in
sharp contrast with Tab.~\ref{fitTauKpi} where the model dependent
uncertainties arising from the $s_{\rm cut}$ dependence were of the
same order as the statistical ones. The central results of
$\lambda_+'$ and $\lambda_+''$ exhibit a small shift because
$K_{l_3}$ experiments favour larger values. The mass of the $K^*(892)$,
in its turn, turns out to be almost the same as in the previous fit
and is still very stable with respect to changes in $s_{\rm cut}$.
The $K^*(892)$ width is slightly larger than before but compatible
within one sigma with the previous result. The parameters of the
second resonance have still large uncertainties but remain compatible
with the results of Tab.~\ref{fitTauKpi}. Finally the normalisation
$\bar B_{K\pi}$ turns out larger than in the previous fit due to a
positive correlation with $\lambda_+'$ and $\lambda_+''$ but
fully compatible with the experimental experimental average $B_{K\pi}^{\rm exp}=0.418(11)$.

%\EPSFIGURE[]{Fig1_MainFit,width=0.75\columnwidth}{Fit result for the spectrum of $\tauKpi$  with $s_{\rm cut}=4$~GeV$^2$, third column of Tab.~\ref{fitCombined}. The data are from the Belle collaboration~\cite{Belle}. Points represented with unfilled circles are excluded from the fit (see text in Section~\ref{Fit1}). The solid red line represents the full fit including contributions from $F_+(s)$ and $F_0(s)$. The scalar contribution alone is represented by the dot-dashed orange line whereas the dashed blue line gives the vector contribution.}

\FIGURE[!ht]{
\includegraphics[width=0.85\columnwidth,angle=0]{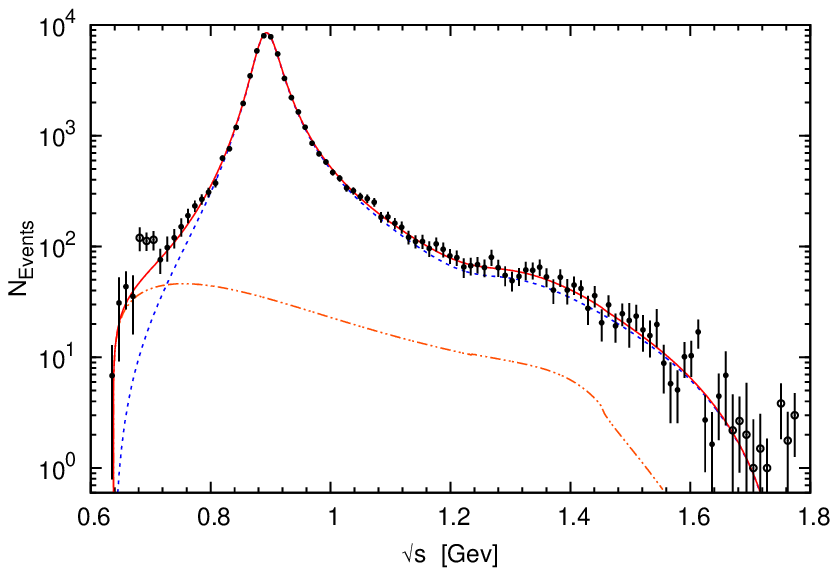}
\caption{{Fit result for the spectrum of $\tauKpi$  with $s_{\rm cut}=4$~GeV$^2$, third column of Tab.~\ref{fitCombined}. The data are from the Belle collaboration~\cite{Belle}. Points represented with unfilled circles are excluded from the fit (see text in Section~\ref{Fit1}). The solid red line represents the full fit including contributions from $F_+(s)$ and $F_0(s)$. The scalar contribution alone is represented by the dot-dashed orange line whereas the dashed blue line gives the vector contribution.  } } \label{MainFitFigure}
}

\TABLE{
\begin{tabular} {|l c c c c c c c c|}
\hline\hline
   & $\bar B_{K\pi}$ &  $m_{K^*}$ & $\Gamma_{K^*}$  &$m_{K^{*\prime}}$ &  $\Gamma_{K^{*\prime}}$  &$\gamma $  &  $\lambda_+^{'} $   &  $\lambda_+^{''}$      \\ 
\hline
%1. &  &  &  &  &  &  &  \\
 $m_{K^*}$&-0.193 & 1 &  &  &  &  &  &  \\
$\Gamma_{K^*}$& -0.414 & -0.007 & 1 &  &  &  &  &  \\
 $m_{K^{*\prime}}$& 0.223 & -0.233 & -0.043 & 1 &  &  &  &  \\
 $\Gamma_{K^{*\prime}}$& -0.261 & 0.243 & 0.130 & -0.675 & 1 &  &  &  \\
$\gamma $ &  0.399 & -0.344 & -0.193 & 0.630 & -0.886 & 1 &  &  \\
 $\lambda_+^{'} $& 0.316 & 0.058 & 0.290 & -0.186 & 0.252 & -0.386 & 1 &  \\
 $\lambda_+^{''}$   &0.776 & -0.233 & 0.045 & -0.001 & 0.054 & 0.006 & 0.747 & 1\\
\hline\hline
\end{tabular}
\caption{Correlation coefficients for the parameters of the fit with  $s_{\rm cut}=4$~GeV$^{2}$, third column of Tab.~\ref{fitCombined}. }
\label{Correlations2}}

%%%%%%%%%%%%%%%%%%%%%% SECTION WITH PHASE SPACE INTEGRALS
\section{\boldmath{$K_{l_3}$ phase space integrals}}
\label{IntKl3}

From the results of our fits shown in Tab.~\ref{fitCombined} one can 
calculate the phase-space integral needed in the computation of $K_{l_{3}}$ decay widths.
The phase-space integral is defined as\footnote{We employ the notation of Ref.~\cite{FlaviaNet} but the definition of the integral is identical to that of Ref.~\cite{LR84}.}
\bea
\label{PhSpaceInt}
 I_{K_{l_{3}}} =  \frac{1}{m_K^2}&&\!\!\! \int\limits_{m_l^2}^{(m_K-m_\pi)^2} \rd t\, \lambda(t)^{3/2} \left( 1+ \frac{m_l^2}{2t}  \right)  \left( 1- \frac{m_l^2}{t}  \right)^2 \times\nn \\
&&\times
\left(  |\tilde F_+(t)|^2 + \frac{3 \, m_l^2 (m_K^2-m_\pi^2)^2}{(2t+m_l^2)\,m_K^4\,\lambda(t)}|\tilde F_0(t)|^2\right)\, ,
\eea
where $m_l$ is the mass of the lepton and
\beq\label{lambda}
\lambda(t)=1 +t^2/m_K^4+ r_\pi^4-2\,r_\pi^2 -2\,r_\pi^2\, t/m_K^2 -2\,t/m_K^2
% lambda of Leutwyler and Roos\lambda(t)=t^2+m_K^4+m_\pi^4 - 2\, t\, m_\pi^2 - 2\, t\, m_K^2 - 2\, m_\pi^2\, m_K^2.
\eeq
with $r_\pi=m_\pi^2/m_K^2$.
In the phase-space integral for the decays with an electron in the
final state, $I_{K_{e_{3}}}$, the smallness of the electron mass makes
the contribution of $F_0$ immaterial. The scalar form factor gives
nevertheless a non-negligible contribution for $K_{\mu_{3}}$
decays. In phase-space integrals for  decays of charged kaons,  we have assumed
that the normalised form factors $\tilde F_{+,0}$ are isospin invariant, which amounts
to assuming that isospin breaking effects are solely contained in $F_+(0)$. Then, for the phase-space
factors of charged-kaon integrals  we employ the mass of the charged kaon and
that of the neutral pion.
 Tab.~\ref{PSInt} contains our results for the integrals. In order to 
take into account all errors and correlations,  a  MC sample of parameter 
values employing the results from Tabs.~\ref{fitCombined} and~\ref{Correlations2}
was generated. The integrals were computed for each set of parameters in these samples.
The errors quoted in  Tab.~\ref{PSInt}  are of a gaussian nature to a good approximation.

\TABLE{
\begin{tabular} {|l c c c c|}
\hline\hline
  &  $s_{\rm cut}=3.24$ GeV$^2$  &  $s_{\rm cut}=4$ GeV$^2$ & $s_{\rm cut}=9$ GeV$^2$  & $s_{\rm cut}\to \infty$  \\ 
  \hline
$I_{K^0_{e_{3}}}$    &   0.15463(17)  &    0.15465(16)      &     0.15468(16) &      0.15468(16)   \\
$I_{K^0_{\mu_{3}}}$   &   0.10275(10)  &    0.10276(10)      &     0.10277(10) &      0.10277(10)   \\
$I_{K^+_{e_{3}}}$    &   0.15900(17)  &    0.15902(16)      &     0.15905(16) &      0.15905(16)   \\
$I_{K^+_{\mu_{3}}}$   &   0.10573(11)  &    0.10575(10)      &     0.10576(10) &      0.10576(10)   \\
\hline\hline
 \end{tabular}
\caption{Results for the phase-space integrals defined in Eq.~(\ref{PhSpaceInt}) obtained with parameters from the fits of  Tab.~\ref{fitCombined}. The uncertainties include the statistical errors and correlations from the fit.}
\label{PSInt}}

%%%%%%%%%%%%%% SEC IV: KPI PHASE SHIFT%%%%%%%%%%%%
\section{\boldmath{$K\pi$} isospin-1/2 $P$-wave scattering phase}
\label{Res}

The decay $\tauKpi$ offers a good environment for the study of $K\pi$
dynamics. From the point of view of strong interactions,  the $K\pi$ pair in the final 
state is isolated.  As a matter of fact, this decay is
certainly a better laboratory for the study of the $K\pi$ phase than
the hadronic reactions used in the classical determinations of the
$K\pi$ phase shifts.  Watson's theorem states that below the first
inelastic threshold the form factors and the respective partial-wave
scattering amplitudes share the same phase~\cite{Watson}.  In the case
of the $P$ wave, the first inelastic channel one can consider is the
quasi-two-body $K^* \pi$ which opens at $\sim 1030$
MeV~\cite{Moussallam}. Therefore, below this value, the phase of our
vector form factor can be compared with the respective scattering results.  In
Fig.~\ref{Phase}, we compare our phase with those from
LASS~\cite{LASS} and Estabrooks {\it et al.}~\cite{Estabrooks}. In this
comparison, one
should bear in mind that isospin breaking effects could play a small role since
the hadronic experiments measured the neutral channel whereas we have
the charged one. Nevertheless, from the inspection of
Fig.~\ref{Phase}, it is clear that our results are compatible with the
experimental determinations of the $K\pi$ $I=1/2$ $P$-wave scattering
phase shift between 850 MeV and  roughly $1$~GeV, just before inelasticity sets in.  From
threshold up to 850 MeV our results seem to be systematically lower
than those from hadronic reactions. It is interesting to remark that
the same behaviour is also observed in the recent Roy-Steiner-type analysis
of $K\pi$ scattering performed by B\"uttiker, Descotes-Genon and
Moussallam~\cite{KpiRoy}. Their phase is also somewhat below the
experimental data up to about 950 MeV.
Finally, we remind that the low-energy results from Estabrooks et
al.~\cite{Estabrooks} have been shown to be inconsistent with a
dispersive analysis of $K\pi$ scattering~\cite{Lang} and,
unfortunately, LASS results~\cite{LASS} do not span the energy region
close to threshold.

\FIGURE[]{
\includegraphics[width=0.85\columnwidth,angle=0]{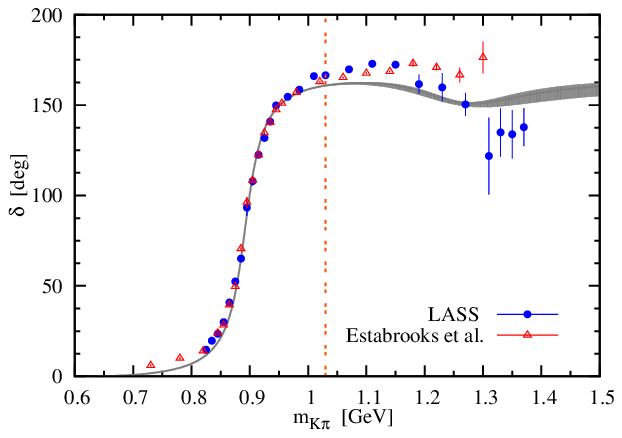}
\caption{{Phase of the form factor $F_+(s)$ together with experimental results  from   LASS~\cite{LASS} and  Estabrooks {\it et al.}~\cite{Estabrooks}. The opening of the first inelastic channel, $K^*\pi$, is indicated by the dashed vertical line. The gray band represents the extrema from the fits of Tab.~\ref{fitCombined}.} }
\label{Phase}
}

Within the elastic domain, the phase of our form factor $F_+(s)$
equals the scattering phase for  the $P$ wave with $I$=1/2.
From the expansion of the corresponding partial wave $t$-matrix near threshold we can
obtain the threshold parameters.  Following  Ref.~\cite{KpiRoy},  they  are defined 
for isospin $I$ and angular momentum $l$ as
\beq
\label{ScattLeng}
\frac{2}{\sqrt{s}}\RE t_l^{I}(s)  = \frac{1}{2 q} \sin 2\delta_l^I (q) = q^{2l} \left[ a_l^I\, +b_l^I\, q^2 + c_l^I\, q^4 +\mathcal{O}(q^6)  \right]\, ,
\eeq
where $q(s)$ is given by Eq.~(\ref{qKpi}). It is simple to express our results in the form of Eq.~(\ref{ScattLeng}) using 
\beq
s= m_K^2 +m_\pi^2 + 2q^2 +2 \sqrt{m_K^2 q^2 + m_\pi^2 q^2 + m_K^2m_\pi^2 + q^4}\, .
\eeq
Then, using Eq.~(\ref{phase}) and  the results of Tab.~\ref{fitCombined} we can compute the
threshold parameters. The first three of them are given in
Tab.~\ref{ScattLengTab} for the four values of $s_{\rm cut}$
investigated in our main fit. The uncertainties in
Tab.~\ref{ScattLengTab} are obtained from a MC that takes into account
all errors and correlations given in Tabs.~\ref{fitCombined}
and~\ref{Correlations2}. One should however note that the functional form of the threshold 
parameters, unlike $\lambda_+'$ and $\lambda_+''$, is determined by our model of $\delta(s)$. Their values  depend mainly upon the masses and widths of the resonances, most notably
that of the $K^*(892)$. Since the pole of the $K^*(892)$ is
very well determined in our fits, the uncertainties in the scattering
lengths are accordingly small.   Tab.~\ref{ScattLengTab} contains only
the propagation of statistical uncertainties. The systematics uncertainty associated with the threshold parameters will be estimated in Sec.~\ref{Conc}.

%\begin{widetext}

\TABLE[!ht]{
\begin{tabular} {|l c c c c|}
\hline\hline
  &  $s_{\rm cut}=3.24$ GeV$^2$  &  $s_{\rm cut}=4$ GeV$^2$ & $s_{\rm cut}=9$ GeV$^2$  & $s_{\rm cut}\to \infty$  \\ 
\hline
$m_{\pi^-}^3 \, a_1^{1/2} \times 10$         &  $0.1658(13)$      &  $0.1656(13)$  &  $0.1655(13)$& $0.1655(13)$  \\
$m_{\pi^-}^5 \, b_1^{1/2} \times 10^{2} $    &  $0.2573(24)$       &   $0.2581(23)$ &  $0.2582(23)$& $0.2583(23)$\\
$m_{\pi^-}^7\,  c_1^{1/2}\times 10^{3}$      & $0.8987(81)$        &   $0.9001(76)$ &   $0.9000(75)$&$0.9000(75)$\\
\hline\hline
  \end{tabular}
\caption{Threshold parameters defined in Eq.~(\ref{ScattLeng}) calculated with the results of our main fit given in Tab.~\ref{fitCombined}. The uncertainties are solely statistical.} 
\label{ScattLengTab}}

%%%%%%%% CONCLUSIONS AND FINAL VALUES %%%%%%%%%
\section{Conclusions}
\label{Conc}

%%% mass and width of the K*(892)

In this section we present our final results. They are obtained from the
main fit displayed in~Tab.~\ref{fitCombined}. Throughout this section,
central values correspond to the average of the extrema found after the variation of
$s_{\rm cut}$ in~Tab.~\ref{fitCombined}.
Let us start with the mass and width of the $K^{*}(892)^\pm$. To the statistical
uncertainty one should add another source of error: the imperfect
knowledge of the detector response.  To that end, we rely on the original analysis performed by the Belle
collaboration where it is found to be 0.44~MeV for the mass of the
$K^*(892)^\pm$ and 1.0~MeV for its width\footnote{The determination of the error due to detector effects is rather involved and depends on the model that is assumed for the analysis. Therefore, we take these values as mere  estimates~\cite{SimonBoris}.}~\cite{Belle}. In principle,
one should include an uncertainty due to the residual dependence on
$s_{\rm cut}$ but Tab.~\ref{fitCombined} shows that the results are
almost invariant under changes of this parameter. Therefore, this
source can safely be neglected. Our final results 
for the mass and width of the $K^*(892)^\pm$ defined from its
pole position as in Eq.~(\ref{pole}) are then
\bea\label{massandwidth}
m_{K^*(892)^\pm}&=& 892.03\pm (0.19)_{\rm stat}\pm (0.44)_{\rm sys}\, \,\mbox{MeV}, \nn\\
\Gamma_{K^*(892)^\pm}&=& 46.53 \pm (0.38)_{\rm stat}\pm (1.0)_{\rm sys}\,\, \mbox{MeV}\, .
\eea
Before comparing this result with  other  analyses of
the same data, one should note that in Refs.~\cite{Belle,JPP2008} a
different {\it definition} of the mass of the $K^*(892)^{\pm}$ was
used.  Therefore,  we have computed the
pole position for the other analyses in order to harmonise the definition of mass.  
Moreover, the use of Eq.~(\ref{pole}) provides   less model dependent results for the resonance parameters~\cite{Escribanoetal}.  
In Fig.~\ref{MassesAndWidths}, we compare the PDG recommended values
$m_{K^*(892)^\pm}^{\rm PDG}=891.66\pm 0.26$~MeV and
$\Gamma_{K^*(892)^\pm}^{\rm PDG} = 50.8\pm 0.9$~MeV~\cite{PDG} with
results for the mass and width of the $K^*(892)^\pm$ obtained from the
{\it pole position} computed from the results of three different
analyses of the Belle data set of $\tauKpi$ decays.  Additional care
should be taken when comparing these results since the PDG values are
obtained chiefly from the parameters of Breit-Wigner-type expressions.
On the basis of our results we claim
that there is no tension between the mass found from $\tau$ decays and
the PDG recommended value {\it provided} the pole position prescription is used for the former.  On
the other hand, the PDG value for the width is only marginally
compatible with the one from Eq.~(\ref{massandwidth}).  The width from
analyses of the Belle data on $\tauKpi$ tend to lower values.
Let us conclude by quoting another unambiguous result that can be derived from our analysis: 
the point $s_{\pi/2}$ satisfying  $\delta(s_{\pi/2})=\pi/2$. Often, this point is used as
the definition of the so-called {\it visible} or  {\it peak} mass of a resonance since it is extracted
from the direct comparison with experimental data\footnote{If the resonance in question is narrow and isolated enough from other resonances and, furthermore, if no background is present, then the mass and width obtained in this way should be equal to the pole mass definition~\cite{Escribanoetal}. We observe that the pole mass is close but not the same as the peak mass.  This is due to the fact that we are not in the ideal situation stated before.}. In our fits, this value
is also very stable with respect to changes in $s_{\rm cut}$ and reads
\beq
\sqrt{s_{\pi/2}}=895.54\pm (0.01)_{s_{\rm cut}}\,\,\,{\rm MeV}\, .
\eeq

\FIGURE[!ht]{
\includegraphics[width=0.48\columnwidth,angle=0]{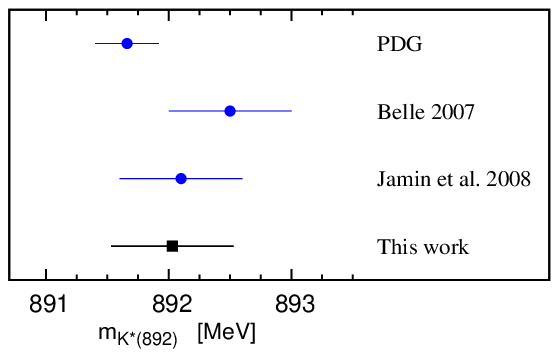}
\includegraphics[width=0.48\columnwidth,angle=0]{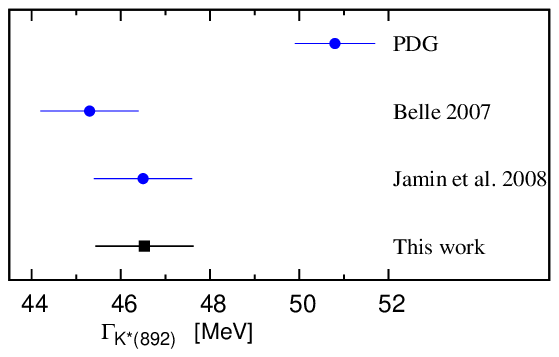}
\caption{{Values for mass and width of the $K^*(892)^\pm$.  On top, we show the PDG recommended values~\cite{PDG}. The other three results are obtained from the  {\it pole position} as in Eq.~(\ref{pole}). The values for pole positions of Belle '07~\cite{Belle} and Jamin {\it et al.} '08~\cite{JPP2008} were computed in Ref.~\cite{BEJ}. To the errors quoted in  Ref.~\cite{BEJ} we have added the systematics uncertainty discussed  in the text.} } 
\label{MassesAndWidths}
}

%%% lambda_+' and lambda_+''

Our final values for $\lambda_+'$ and $\lambda_+''$ come from the
fits of Tab.~\ref{fitCombined}. The results are again very stable with
respect to changes in $s_{\rm cut}$. However, since now the statistical
uncertainties are quite small, this model dependence  
contributes to the total error (specially in the case of $\lambda_+''$). From the mean
of values of  Tab.~\ref{fitCombined} we obtain
\bea\label{lambdas}
\lambda_+'\times 10^{3}  &=&  25.49\pm (0.30)_{\rm stat} \pm (0.06)_{s_{\rm cut}}\,, \nn \\
\lambda_+''\times 10^{4} &=&  12.22\pm (0.10)_{\rm stat} \pm (0.10)_{s_{\rm cut}}\,.
\eea
Concerning $\lambda_+'$, Fig.~\ref{lambdap} shows that the results
from $K_{l_{3}}$ and $\tauKpi$ decays are in very good agreement\footnote{For consistency we
compare our results with dispersive analyses of $K_{l_3}$ decays. Data analyses
that employ the quadratic Taylor expansion  of Eq.~(\ref{FFTaylor}) have much larger errors but agree as well
with our numbers. For  results from the quadratic form factor, see for instance~\cite{ISTRAquad} (ISTRA+),~\cite{KLOE} (KLOE),~\cite{NA48quad} (NA48), and~\cite{KTeVquad} (KTeV).}.  Our
combined analysis produces a result in agreement with the others and
with a rather small uncertainty. For $\lambda_+''$ the situation is
somewhat different. Due to the restricted phase-space, quadratic fits
of $K_{l_3}$ data do not provide a good determination of $\lambda_+''$
and dispersive analyses employ form factors with two subtractions,
hence with only one subtraction constant determined directly from the
data, namely $\lambda_+'$.  In Fig.~\ref{lambdap}, for $K_{l_3}$
experiments, we display results derived from the two-times subtracted
form factor of Ref.~\cite{Bernard:2009zm}.  We compare our results
also to an average of analyses that employ Eq.~(\ref{FFTaylor}) for
$F_+$~\cite{Antonelli10}. Results from $\tau$ decay data have a 
better precision, and are compatible with results from $K_{l_3}$
experiments within their larger error bands. 

From the expansion of Eq~(\ref{dispFF}) we can calculate the  third coefficient of a Taylor series of the type of Eq.~(\ref{FFTaylor}) as
\beq
\lambda_+'''= \alpha_1^3+ 3\, \alpha_1\, \alpha_2 +m_{\pi^-}^6\,\frac{6}{\pi}\int\limits_{s_{K\pi}}^{s_{\rm cut}} ds' \, \frac{\delta(s')}{(s')^4}\, .
\eeq
Then, from the results of our fits, we find for $\lambda_+'''$
\beq
\lambda_+'''\times 10^5=8.87\pm (0.08)_{\rm stat} \pm (0.05)_{s_{\rm cut}}\,,
\eeq
which is again compatible with the corresponding result of Eq.~(5.7) of Ref.~\cite{BEJ}.

The ChPT expansion of $F_+(q^2)$  at $\mathcal{O}(p^4)$ is governed by the low-energy constant
$L_9^r$. Therefore, at this order, from our value of $\lambda_+'$ we can obtain
 $L_9^r$.  It is not our aim here to carefully determine  $L_9^r$, but it
is certainly interesting to check the consistency of our results with the chiral expansion of $F_+$.
Using the  $\mathcal{O}(p^4)$ expressions of Ref.~\cite{GL} with $F_0^2=F_\pi^2$ we obtain
\beq
L_9^r(m_{K^*})\big|_{F_0^2=F_\pi^2}\times 10^{3} = 5.19\pm (0.07)_{\rm stat}\, .
\eeq
It is however well known that the dominant uncertainty is given by the truncation of the series
at  $\mathcal{O}(p^4)$. As an estimate of $\mathcal{O}(p^6)$ effects we can employ $F_0^2=F_\pi F_K$
which gives
\beq
L_9^r(m_{K^*})\big|_{F_0^2=F_\pi F_K}\times 10^{3} = 6.29\pm (0.08)_{\rm stat}\, .
\eeq
Our results agree with the one  obtained in Ref.~\cite{BijnensTalavera} from the pion electromagnetic form
factor using  $\mathcal{O}(p^6)$ results: $L_9^r(m_{K^*})\times 10^{3} = 5.70\pm 0.43$.

\FIGURE[!ht]{
\includegraphics[width=0.48\columnwidth,angle=0]{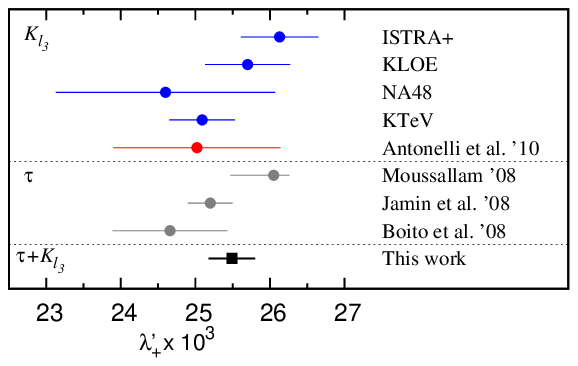}
\includegraphics[width=0.48\columnwidth,angle=0]{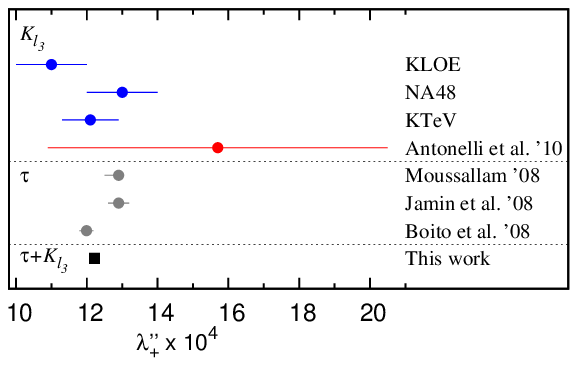}
\caption{{Results for $\lambda_+'$ (left-hand panel) and  $\lambda_+''$ (right-hand panel). The first four come from the dispersive representation of Ref.~\cite{Bernard:2009zm} fitted to $K_{l_3}$ data of~ISTRA+~\cite{Istra},   KTeV~\cite{KTeV}, KLOE~\cite{KLOE}, and NA48~\cite{NA48}.  $\lambda_+''$ of  KLOE and  NA48 are derived from their analyses in Ref.~\cite{Passemar}. The results from Antonelli {\it et al.} '10.~\cite{Antonelli10} are averages of fits using a quadratic form factor [see Eq. (\ref{FFTaylor})]. The central portions of both panels  display results from Moussallam '08~\cite{Moussallam}, Jamin {\it et al.} '08~\cite{JPP2008}, and our previous analysis~\cite{BEJ}.}}\label{lambdap}}

 %%% Integrals IKl3

Our results for the phase space integrals of $K_{l_{3}}$ decays have
been collected in Tab.~\ref{PSInt}. Following the procedure outlined
above we find the final results given in the second column of
Tab.~\ref{PhaseSpaceInt}. In the same table, we display the results of
the compendium performed in Ref.~\cite{Antonelli10} from dispersive
and quadratic fits to $K_{l_{3}}$. Our results are compatible with
those found in Ref.~\cite{Antonelli10}.

\TABLE{
\begin{tabular} {|l c c c |}
\hline\hline
                   & This Work &  $K_{l_{3}}$  disp.~\cite{Antonelli10} & $K_{l_{3}}$ quad.~\cite{Antonelli10}  \\
 \hline
$I_{K^0_{e_{3}}}$    &   0.15466(17)    &     0.15476(18) &    0.15457(20)       \\
$I_{K^0_{\mu_{3}}}$   &   0.10276(10)    &     0.10253(16) &    0.10266(20)        \\
$I_{K^+_{e_{3}}}$    &   0.15903(17)     &     0.15922(18) &    0.15894(21)        \\
$I_{K^+_{\mu_{3}}}$   &   0.10575(11)    &     0.10559(17) &    0.10564(20)      \\
\hline\hline
  \end{tabular}
\caption{Results for the $K_{l_{3}}$ phase-space integrals. Our results include a small uncertainty due to the dependence on $s_{\rm cut}$.  For comparison,
we also give the results of Ref.~\cite{Antonelli10} that come from averages of quadratic (quad.) and  dispersive (disp.) analyses of  $K_{l_{3}}$ data.    } 
\label{PhaseSpaceInt}
}

%%%%%%%%%%%% Scattering lengths    %%%%%%%%%%%%%%%%%

Finally, in Tab.~\ref{ScattLengTabComp} we present our final values
for the $K\pi$ $P$-wave $I$=1/2 threshold parameters. These results are
compared with other results found in the literature. Our final numbers
include the statistical uncertainty as well as the (small) $s_{\rm
cut}$ dependence added in quadrature.  Furthermore, we propagate the
additional error of Eq.~(\ref{massandwidth}) in order to account for
systematics. However, the precision obtained for the $K^*(892)$ pole
is such that our values have smaller uncertainties as compared to
other determinations of the threshold parameters. The main discrepancy
observed is in the value of the effective range $b_1^{1/2}$ that turns out substantially
larger than that of Ref.~\cite{KpiRoy}. In that reference, however,
the authors already noted that their results could be affected by
the uncertainties of LASS~\cite{LASS} data at energies above $1$~GeV. The point where
their phase equals $\pi/2$ is also shifted by $10$~MeV as compared to
ours. Therefore, since our data set is not contaminated with spurious
strong interactions in the final state, we consider this discrepancy to
be harmless.

\TABLE[!h]{
\begin{tabular} {|l c c c c c|}
\hline\hline
  &  This work  &    \cite{KpiBKM} &  \cite{Bijnens}& \cite{RChPTKpiBKM} &  \cite{KpiRoy}  \\ 
\hline
$m_{\pi^-}^3 \, a_1^{1/2} \times 10$          &  0.166(4)   &    0.16(3)  &   0.18    &   0.18(3)      &    0.19(1)\\
$m_{\pi^-}^5 \, b_1^{1/2} \times 10^{2} $     &  0.258(9)   &    -        &      -       &    -       &    0.18(2)\\
$m_{\pi^-}^7\,  c_1^{1/2}\times 10^{3}$       &  0.90(3)   &    -        &      -         &    -       &    0.71(11)\\
\hline\hline
  \end{tabular}
\caption{Our final values for the threshold parameters compared with results found in the literature. In Ref.~\cite{KpiBKM}  ChPT at $\mathcal{O}(p^4)$ was used whereas in Ref.~\cite{Bijnens} ChPT at $\mathcal{O}(p^6)$ was employed. Results from Ref.~\cite{RChPTKpiBKM} are obtained within RChPT at $\mathcal{O}(p^4)$ and in Ref.~\cite{KpiRoy} a Roy-Steiner dispersive analysis of $K\pi$ scattering was carried out.   } 
\label{ScattLengTabComp}}

A final point concerning $K\pi$ interactions that should be address is
the existence of the controversial low-mass $S$-wave isospin-$1/2$ resonance
$K^*_0(800)$ (or simply $\kappa$). In the description of the scalar
form factor used here~\cite{JOP4}, a pole that can be identified with
the $\kappa$ is present on the second Riemann sheet of the corresponding
scattering amplitude~\cite{JOPScatt}. Therefore, the success of our description
of the spectrum in the low-energy region corroborates the existence of such a
state.

%%%%%%%%%%% FINAL PARAGRAPH
In conclusion, dispersion relations provide a technique to construct form factors
valid for the description of $\tau$ and kaon decay data. In the light
of our results, we are confident that the use of dispersive form factors to fit
the spectrum of $\tauKpi$ with restrictions from $K_{l_3}$ experiments
is a valid strategy towards the improvement of our knowledge of $K\pi$
form factors. Furthermore, some aspects of $K\pi$ dynamics can also be probed.
New results for the spectrum of $\tauKpi$ from other collaborations
would offer a very good prospect to further improve our analysis.
%The results could be improved further by means of a joint 
%analysis of kaon and tau data.

%%%%%%%%%%%%%% ACknowledgments %%%%%%%%%%%%
\acknowledgments
We are grateful to the Belle collaboration in particular to S.~Eidelman,
D.~Epifanov and B.~Shwartz, for providing their data and for useful
discussions. We also benefited from discussions with B.~Ananthanarayan, M.~Antonelli,
V.~Bernard, J.~Bijnens, B.~Moussallam, and E.~Passemar. DRB thanks
B. Moussallam and the hospitality of the {\it Institut de Physique
Nucl\'eaire} at {\it Universit\'e Paris XI} where this work was
completed.  This work was supported in part by the Ministerio de
Ciencia e Innovaci\'on under grant CICYT-FEDER-FPA2008-01430, the EU
Contract No.~MRTN-CT-2006-035482, ``FLAVIAnet'', the Spanish
Consolider-Ingenio 2010 Programme CPAN (CSD2007-00042), and the
Generalitat de Catalunya under grant SGR2009-00894. We also thank the {\it Universitat
Auton\`oma de Barcelona.}

%%%%%%%%             APPENDIX        %%%%%%%%%
\appendix
\section{Form factors}
\label{AppA}

\subsection{Vector form factor}
For the phase $\delta(s)$ needed in order to employ Eq.~(\ref{dispFF}) we take a form inspired by the  RChT treatment of~Refs.~\cite{JPP2006,JPP2008}
with two vector resonances. As described in greater detail in Ref.~\cite{BEJ},
$\delta(s)$ can be cast into the following form
\beq\label{phase}
\delta(s) = \tan^{-1}\left[\frac{\IM\, \tilde f_+(s)}{\RE\, \tilde  f_+(s)}\right]\, ,
\eeq
where
\begin{equation}
\label{FpKpi2}
\tilde f_+(s) \,=\, \frac{\tilde m_{K^*}^2 - \kappa_{K^*}\,\tilde H_{K\pi}(0) +
\gamma \, s}{D(\tilde m_{K^*},\gamma_{K^*})} -
\frac{\gamma\, s}{D(\tilde m_{K^{*'}},\gamma_{K^{*'}})} \,.
\end{equation}
The first term in the right-hand side of Eq.~(\ref{FpKpi2})
corresponds to the $K^*(892)$ whereas the second represents the
contribution of the second vector resonance $K^*(1410)$. The mixing
parameter $\gamma$ is obtained from the fits and $\tilde H_{K\pi}(s)$
is the one-loop $K\pi$ bubble integral, whose precise definition is given
in Refs.~\cite{JPP2006,GL}. The denominators $D(\tilde m_{K^*}, \gamma_{K^*})$ are
\begin{equation}
\label{Den}
D(\tilde m_n,\gamma_n) \,\equiv\, \tilde m_n^2 - s - \kappa_n \, \RE\,\tilde H_{K\pi}(s) -
i\, \tilde m_n \gamma_n(s) \,,
\end{equation}
where the constants 
\beq
\kappa_n= \frac{192 \pi F_K F_\pi}{\sigma(\tilde m_n^2)^3}\frac{\gamma_n}{\tilde m_n}
\eeq
 are defined so that $-i \kappa_n\, \IM\, \tilde H_{K\pi}(s) = - i \tilde m_n \gamma_n(s)$ and the running width of a vector resonance is taken to be 
\beq\label{width}
\gamma_n(s) = \, \gamma_{n}\frac{s}{\tilde m_{n}^2} \frac{\sigma^3_{K\pi}(s)}
{\sigma^3_{K\pi}(\tilde m_{n}^2)} \,.
\eeq
The phase-space function $\sigma_{K\pi}(s)$ is given by
$\sigma_{K\pi}(s)= 2 \, q_{K\pi}(s)/\sqrt{s}$, whereas $q_{K\pi}$ is
defined in Eq.~(\ref{qKpi}). The model parameters $\tilde m_n$ and
$\gamma_n$ are not the physical resonance mass and width. Physical
values are obtained solving the equation $D(\tilde m_n, \gamma_n)=0$ for
complex values of $s$.  Consequently, our definition of physical mass and
width is given by Eq.~(\ref{pole}).

\subsection{Scalar form factor}

 The procedure adopted in Ref.~\cite{JOP1}
is to solve the multi-channel Muskelishivili-Omn\`es problem for 3
channels (where $1\equiv K\pi, 2\equiv K\eta$ and $3\equiv K
\eta'$). Each of the scalar form factors $F_0^k$, where $k$ represents the channel, is then coupled to
the others via
\beq
F_0^k(s)= \frac{1}{\pi} \sum_{j=1}^3\,  \int\limits_{s_j}^{\infty} d s'\, \frac{\sigma_j(s')F_0^j(s')t_0^{{k\to j}}(s')^*}{(s' -s -i\epsilon)}\, . \label{F0}
\eeq
In the last equation, $s_j$ is the threshold for channel $j$, $\sigma_j$(s)
are two-body phase-space factors and $t_0^{k\to j}$ are partial wave
$T$-matrix elements for the scattering $k \to j$. The form factors are
obtained solving the coupled dispersion relations arising from
Eq.~(\ref{F0}). This is done imposing chiral symmetry constraints and using
$T$-matrix elements from Ref.~\cite{JOPScatt} that provide a good description of
scattering data. Within the elastic approximation, Eq.~(\ref{F0}) reduces to
the usual single-channel Omn\`es equation.

\end{document}